\documentclass{PoS}
\usepackage{amssymb,amsmath,array}

\title{Self-Orgazing Maps Parametrization of Parton Distribution Functions}

\ShortTitle{Self-Orgazing Maps Parametrization of Parton Distribution Functions}
 
\author{Daniel Z. Perry \\ University of Virginia - Physics Department \\
382, McCormick Rd., Charlottesville, Virginia 22904 - USA \\ 
E-mail: \email{dzp3h@virginia.edu}} 

\author{Katherine Holcomb \\ 
University of Virginia - UVACSE \\  
112, Albert H. Small Building, Charlottesville, Virginia 22904 - USA \\ 
E-mail: \email{kholcomb@virginia.edu}} 

\author{Simonetta Liuti \\ University of Virginia - Physics Department \\
382, McCormick Rd., Charlottesville, Virginia 22904 - USA \\ 
E-mail: \email{sl4y@virginia.edu}}
\abstract{We describe a new method to extract parton distribution functions both in the unpolarized and the
polarized case, based on a type of neural networks, the Self-Organizing Maps. Initial quantitative results of our Next to Leading Order analysis are presented for the unpolarized case.}

\FullConference{XVIII International Workshop on Deep-Inelastic Scattering and Related Subjects\\
                April 19 -23, 2010\\
                Convitto della Calza, Firenze, Italy}

\begin{document}
\section{Introduction}
Deep inelastic processes  provide our primary evidence for studying the internal structure of hadrons in terms of quarks and gluons as predicted from QCD.
The factorization properties of QCD allow us to write the cross sections by separating the hard part, which is the well defined interaction of a quark with {\it e.g.} a virtual photon, from the soft part, described in terms of Parton Distributions Functions (PDFs). The latter cannot be calculated from first principles;
thus information about the PDFs can only be obtained directly from experiment. 
The observables obtained from experiment require models/analytic forms in order to describe and understand their behavior. Several collaborative groups have sought to use fitting techniques to constrain these forms and to extract the PDFs, $f_i(x,Q^2)$, $i=u,d,s.,c,b,t,g$, with light cone momentum fraction $x$, at a scale $Q^2$.
The fits results vary significantly from group to group and there is no well defined theory to constrain them (for a review of recent results see {\it e.g.} \cite{Dittmar} and \cite{PDF_WG}).  

On the other side, a relatively new approach to PDFs fitting is the one proposed by
the NNPDF collaboration \cite{Rojo},
who have replaced the standard analytic forms with 
a more complex  Neural Network (NN) solution, and the Hessian method
with Monte Carlo (MC) sampling of the data.
The NNPDF method also circumvents the problem inherent in global fitting, of choosing a suitable 
$\Delta \chi^2$.
The estimated uncertainties for NNPDF fits are larger than those of global
fits, possibly indicating that the global fit uncertainties might be underestimated.

In \cite{Carnahan} a criticism was put forward about relying on purely automated fitting procedures such as the ones used by NNPDF. A new specific type of neural network, the Self-Organizing Map (SOM), was proposed. 
The main point is that since for NNPDFs the effect of modifying individual
NN parameters is unknown, the result might not be under control
in the extrapolation region, or in between the data points if the data are 
sparse. In this case, as well as in the case of incompatible data, the NNPDF method might 
be doubtful.

In the new LHC era it is of the utmost importance to provide methods to carefully extract PDFs and their uncertainties  in order to  interpret the soon-available increasingly complicated and diverse 
sets of observations. 
Motivated by this, and also by the possibility of extending our program to a wider set of semi-inclusive and exclusive observables, we pushed forward with the SOM method, and we improved the preliminary work in \cite{Carnahan} by restructuring our original code in such a way that we obtain smooth, continuous types of solutions for which a fully quantitative error analysis can be implemented.  Our new code is now sufficiently flexible to allow for analyses of different observables, from both the unpolarized and polarized inclusive type structure functions, to the matrix elements for deeply virtual exclusive and semi-inclusive processes.
Our first quantitative results for the unpolarized case using Next-to-Leading-Order (NLO) perturbative QCD in the $\overline{MS}$ scheme are shown in this workshop. 

\section{Self-Organizing Maps}
\label{sec2}
SOMs were developed by T. Kohonen in the '80s \cite{Kohonen}. They are a type of neural network whose nodes/neurons -- map cells -- are tuned to a set of input signals/data/samples according to a form of adaptation. The various nodes form a topologically ordered map during the learning process. This feature constitues perhaps the main strength of SOMs in that it allows one to obtain a simplified two-dimensional representation of complex data otherwise depending on a "hard to control''  set of parameters. 
What sets SOMs apart from other NNs is also their learning process. SOMs learn via {\em unsupervised learning} whereas the learning 
of generic artificial NNs is {\em supervised}. 
In supervised learning a set of examples is given, and the goal is to force the data to match the examples as closely as possible.  A cost function is defined that measures the importance to miss or detect the correct result. During the learning process the cost function is minimized. In unsupervised learning the cost function is minimized without introducing a definite set of examples, but just by similarity relations, or  by finding how the data cluster or self-organize.  Many new 
uses described below might derive from this property.    

\subsection{SOM Algorithm}
\label{sec2.1}
The SOM algorithm consists of three stages: {\it i)} Initialization; {\it ii)} Training; {\it iii)} Mapping. 
Each cell (neuron) is sensitized to a different domain of vectors.
 
During the initialization procedure weight vectors of dimension $n$ are associated to each cell $i$:
\[ V_i = [v_i^{(1)}, ..., v_i^{(n)} ]  \] 
$V_i$ are given spatial coordinates, {\it i.e.} one defines the geometry/topology of a 2D map
that gets populated randomly with $V_i$. 
For the training, a set of input data
\[ \xi = [\xi_i^{(1)}, ..., \xi_i^{(n)} ] , \] 
(isomorphic to $V_i$) is then presented  to $V_i$, or compared  via a "similarity metric" that we choose to be
\[ L_2(x,y) = \sum_{i=1,2} \sqrt{  x^2_i - y^2_i  } \]
The most similar weight vector is the Best Matching Unit (BMU). SOMs are based on  unsupervised and "competitive" learning.
This means that the cells that are closest to the BMU activate each other in order to "learn" from $\xi$.  Practically, they adjust their
values according to
\begin{equation}
V_i(n+1) = V_i(n) +  h_{ci}(n) [\xi(n) - V_i(n)],
\label{learn_1}
\end{equation}
where $n$ is the iteration number, and $h_{ci(n)}$ is the "neighborhood function" defining a radius on the map which decrease with both $n$, and
the distance between the BMU and node $i$. In our case we use square maps of size $L_{MAP}$, and 
\begin{equation}
h_{ci}(n) = 1.5 \left(\frac{n_{train} -n}{n_{train}} \right) L_{MAP}
\label{learn_2}
\end{equation}
where $n_{train} $ is the number of iterations.
At the end of a properly trained SOM, cells  that are topologically close to 
each other will contain data which are similar to each other.
In the final phase the actual data gets distributed on the map and 
clusters emerge. 

Since each map vector now
represent a class of similar objects, the SOM
is an ideal tool to
visualize high-dimensional data, by projecting it onto a low-dimensional map
clustered according to some desired similar feature. 
In what follows we apply these ideas to PDF fitting.

\section{SOMPDF Parametrization} 
\label{sec3}
We used SOMs to generate and classify possible ``candidate" distribution functions that fit a subset of all available DIS data. 
In a nutshell, a set of database/input PDFs is formed by selecting at random from a range of existing PDF sets and varying their parameters. Baryon number and momentum sum rules are imposed at every step. These input PDFs, evolved to NLO to the desired $Q^2$,  are used to initialize the map (Section \ref{sec2.1}). A subset of input PDFs is then used to train the map. 
The similarity is tested by comparing the PDFs, according to Eqs.(\ref{learn_1},\ref{learn_2}) 
at given $x$ and $Q^2$ values. The new map PDFs are obtained by averaging the neighboring PDFs with the BMU PDFs. Once the map is trained a Genetic Algorithm (GA) is implemented. The $\chi^2$ per input PDFs is calculated with respect to the experimental data. We then take a subset of these functions with the best  $\chi^2$, and use them as seeds to form a new set of input PDFs. We train the map with the new set of input PDFs and repeat the process. The $\chi^2$ was found to decrease monotonically towards $\chi^2=1$ with every GA iteration.  Our stopping criterion is established when the $\chi^2$ stops varying (its curve flattens).
\begin{figure}
\centering
\centerline{\includegraphics[width=0.75\textwidth]{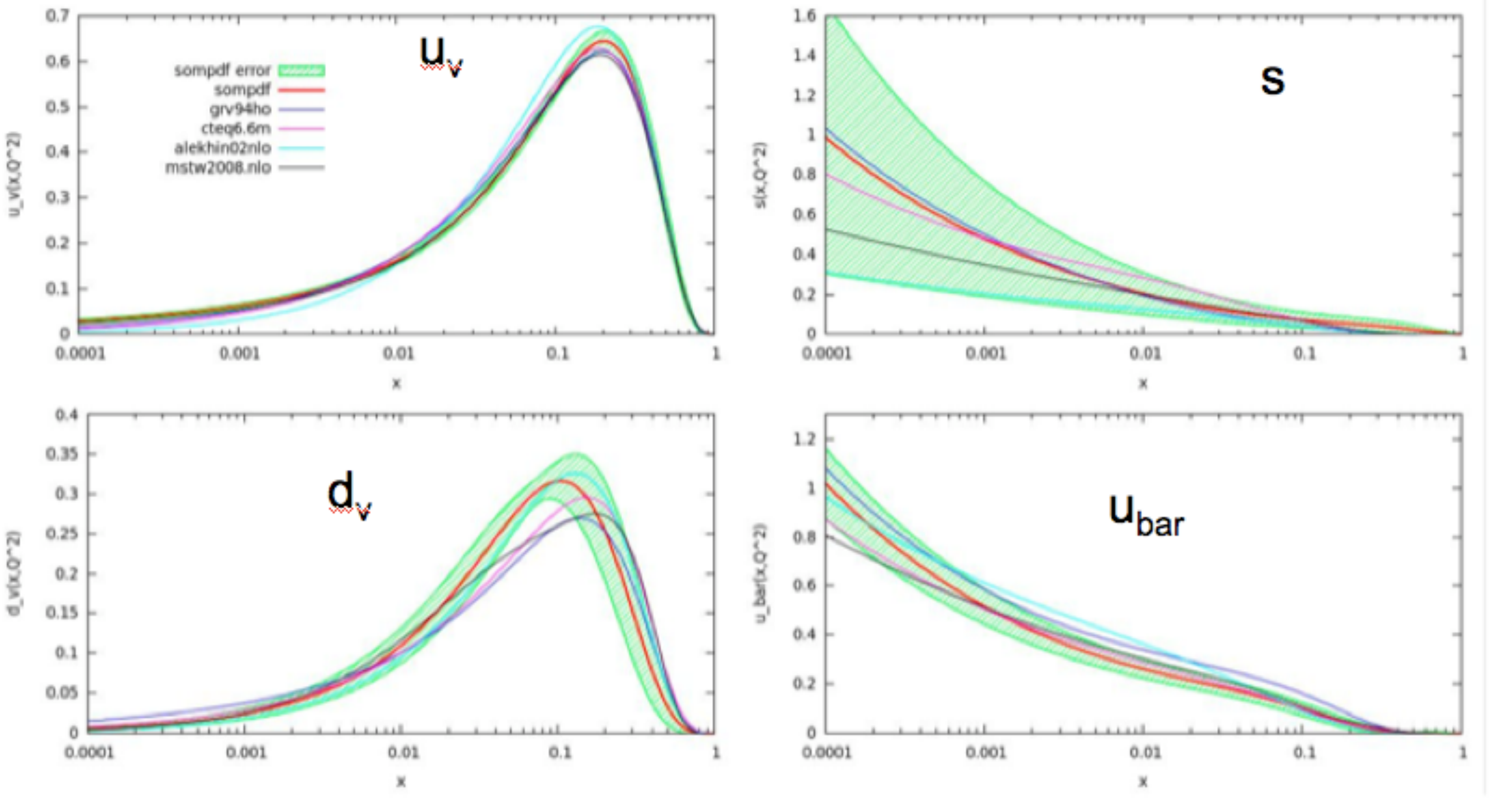}}
\caption{Test results using a $5 \times 5$ map for a set of 43 runs. The panels represent (clockwise from upper left)the $u_v$,
$s$, $d_v$, $\bar{u}$ distributions, respectively, at $Q^2 = 7.5$ GeV$^2$. The shaded areas are our results including the error analys
outlined in the text. For comparison we show also results from several other parametrizations.}
	\label{fig1}
\end{figure}
Our first runs, presented in Figure \ref{fig1}, used a "test" set of data from DIS  consistent with the sets used in Refs. \cite{NNPDF,CTEQ6,AMP06}.  The data sets chosen were from 
BCDMS, H1, NMC, SLAC and ZEUS (see {\it e.g.} references in \cite{NNPDF}). 

As recently articulated in \cite{CTEQ_2010}, the evaluation of the PDFs' uncertainty is complicated because it originates from different experimental and theoretical sources. On one side one deals with the incompatibility of various experimental $\chi^2$'s.On the other, the treatment of theoretical errors is complicated because these cannot be precisely defined, 
and their correlations are not well known.In our approach we defined a statistical error on an ensemble of SOMPDF runs.A more refined evaluation using the Lagrange multiplier method is in progress \cite{progress}.

\section{Conclusions}
In this work we described a new computational method based on Self-Organizing Maps for parametrizing nucleon PDFs. In Ref.\cite{Carnahan} it was shown that this method works well as a minimization technique. However problems connected with the smoothing of the functions remained, and prevented a fully quantitative error analysis.
A solution to this problem was introduced here, obtained from a new version of our code where thanks to its flexibility and to the increased speed of its parallel (MPI) version, we could perform random variations on the parameters of the various input PDFs  thus providing continuous solutions which are amenable to standard error analyses. Applications to polarized structure functions are on their way.

Future developments will also be directed at exploiting the full potential of SOMs that  offer the capability of going beyond a fully automated procedure, by enabling one  to control the fitting procedure at each step. 
The selection of the best PDF candidates for the subsequent iteration could 
then be made based on the user's preferences instead of solely based on the
 $\chi^2$.
Our program can be extended
to multivariable cases such as the Generalized Parton
Distributions where the data is too
sparse for stochastically generated, parameter-free, PDFs.

\vspace{0.3cm} 
We thank the University of Virginia Alliance for Computational Science and Engineering for computer time, and the HPC group at Jefferson Lab, in particular David Richards and Chip Watson, for
allotting us space on their clusters. This work is partially supported by the U.S. Department
of Energy grant DE-FG02-01ER4120 (S.L and D.Z.P.).

\section{Bibliography}
\begin{footnotesize}

\end{footnotesize}


\begin{thebibliography}{99}




\bibitem{url} Slides: \\ 
\verb$http://indico.cern.ch/contributionDisplay.py?sessionId=9&contribId=215&confId=86184$

\bibitem{Dittmar}  A.~De Roeck {\it et al.},
  Eur.\ Phys.\ J.\  C {\bf 66}, 525 (2010)

\bibitem{PDF_WG} R. Placakyte, S. Alekhin, D. Colferai and J.W. Huston, {\it these proceedings}.

\bibitem{Rojo} J. Rojo, {\it these proceedings}; 
\verb$hhtp://sophia.ecm.ub.es/nnpdf/$.
    
\bibitem{Carnahan} H. Honkanen, S. Liuti, J. Carnahan, Y. Loitiere, P. R. Reynolds, Phys. Rev. D 79, 034022 (2009)

\bibitem{Kohonen} T. Kohonen, Self-organizing Maps (Springer, New York, 2001), 3rd. ed.

\bibitem{NNPDF} R. D. Ball, L. Del Debbio, S. Forte, A. Guffanti, J. I. Latorre, A. Piccione, J. Rojo, M. Ubiali (NNPDF Collaboration), Nucl. Phys. B 809, 1-63 (2009)
\bibitem{CTEQ6} J. Pumplin, D. R. Stump, J. Huston, H. L. Lai, P. Nadolsky, W. K. Tung (CTEQ6), Phys.\ Rev.\  D {\bf 65}, 014013 (2001)
 
\bibitem{AMP06} S. Alekhin, K. Melnikov, F. Petriello (AMP06), Phys. Rev. D 74, 054033 (2006) 

\bibitem{CTEQ_2010} H.~L.~Lai, J.~Huston, Z.~Li, P.~Nadolsky, J.~Pumplin, D.~Stump and C.~P.~Yuan,
  arXiv:1004.4624 [hep-ph].  

\bibitem{progress} K. Holcomb, S.Liuti, D.Z. Perry, and S.K. Taneja, {\it in preparation}.

\end{thebibliography}
\end{document}